\documentclass[prb,
superscriptaddress,showpacs,amsmath,amssymb]{revtex4}

\begin{document}

%devoted to memory of Alexei Alexandrovich Starobinsky

\author{G.E.~Volovik}
\affiliation{Low Temperature Laboratory, Aalto University,  P.O. Box 15100, FI-00076 Aalto, Finland}
\affiliation{Landau Institute for Theoretical Physics, acad. Semyonov av., 1a, 142432,
Chernogolovka, Russia}

\title{Schwinger vs Unruh}

\date{\today}

\begin{abstract}
devoted to memory of Alexei Alexandrovich Starobinsky
\\ \,
\\
  It is shown that the temperatures which characterise the Unruh effect, the Gibbons-Hawking radiation from the de Sitter cosmological horizon and the Hawking radiation from the black hole horizon acquire the extra factor 2 compared with their traditional values. The reason for that is the coherence of different processes. The combination of the coherent processes also allows us to make the connection between the Schwinger pair production and the Unruh effect.
\end{abstract}

\maketitle 
 
 %devoted to memory of Alexei Alexandrovich Starobinsky
 
 \tableofcontents
 
\section{Introduction}

There were many discussions concerning the problem with a factor 2 in the temperature of Hawking radiation.\cite{Akhmedov2006,Akhmedov2007,Akhmedov2008,Akhmedova2008,Akhmedova2009,Gill2010}
 The doubling of the Gibbons-Hawking temperature was also discussed for the de Sitter expansion.\cite{Volovik2009,Volovik2024} Since the Schwinger pair creation bears some features of the thermal radiation, one may also expect the factor 2 problem.

 The Schwinger pair creation\cite{Schwinger1951,Schwinger1953} of particles with mass $M$ and charges $\pm q$ in electric field ${\cal E}$ per unit volume per unit time is given by:
\begin{equation}
\Gamma^{\rm Schw}(M) = \frac{dW^{\rm Schw}}{dt} = 
 \frac{q^2{\cal E}^2}{(2\pi)^3}  \exp{\left(- \frac{\pi  M^2}{q\cal E} \right)} \,.
\label{SchwingePairCreation}
\end{equation}
Since $a=q{\cal E}/M$ corresponds to the acceleration of a charged particle, there were attempts to connect the Schwinger mechanism with the Unruh effect\cite{Unruh1976}, see Refs.\cite{Parentani1997,KharzeevTuchin2005,Kempf2021,Zabrodin2022} and references therein. 

In December 2020 I got from Alexei Starobinsky the following message:

"By the way, I discovered a new wrong result not mentioned in literature,
probably because it is wrong. The result is the following: pair creation
in the constant electric field can be explained by the Unruh effect,
namely, thermal activation by the Unruh temperature. Indeed, let us write
the exponent for the probability $-\pi M^2/q{\cal E}$ as $- 2M/T$ where $T$ is some
"temperature", supposedly the Unruh one. However, it follows immediately
that $T= 2q{\cal E}/\pi M = 2a/\pi$, where $a$ is the acceleration, that is 4 times
larger than the Unruh temperature $a/2\pi$. This 4 times difference shows that the
proposed explanation is wrong."

My response was the following:

"Dear Alexei, I also looked at that. But I got the factor 2 instead of 4.
Since you consider acceleration of one particle, the exponent is $-M/T$, and thus $T=a/\pi$.
Anyway, this is also wrong."

Of course, the similarity between the equations for probabilities suggests that there is some analogy between the Schwinger and Unruh effects, although the factors 4 or 2 come as the serious problems. 
Here we consider these problems using the experience\cite{Volovik2024} with the doubling of the Gibbons-Hawking temperature in case of the cosmological horizon.

If one tries to make the direct analogy between these processes, this is already problematic. The original state is the vacuum in the constant electric field. Being the vacuum it does not provide any physical acceleration. Acceleration in electric field appears only in the presence of a charge particle. That is why one can try to find the situation, when the two effects are physically connected. The connection may arise if we split the pair creation in several steps. In the first step the pair of particles is created by Schwinger mechanism and then in the further steps the Unruh process enters, which is caused by the acceleration of the created charged particles. These quantum processes should take place in unison, i.e. coherently. This means that in the quantum tunneling picture\cite{Volovik1999,Wilczek2000} the corresponding exponents are multiplied.

We consider here such scenario, which allows us to connect the acceleration caused by the electric field with Unruh process. Instead of the factor 4 discussed in the Starobinsky message we obtain the factor 2,  i.e. we come to the same factor 2 problem as in the Hawking radiation from the black hole and in the de Sitter spacetime. However, we show here that in such Schwinger-Unruh connection the factor 2 is not wrong. This factor 2 is natural for the Unruh effect in the same way as the temperature of the de Sitter state is naturally twice the Gibbons-Hawking temperature. The common reason, which provides the factor 2 in the de Sitter expansion and in Schwinger-Unruh process, is that in these two situations there is the coherence between several processes.

The similar arguments for the doubling of temperature can be applied to the pure Unruh effect, and also
to the Hawking radiation from the black hole, where according to 't Hooft, the type of the quantum coherence leads to the doubling of the Hawking temperature.\cite{Hooft2022,Hooft2023}  

\section{Schwinger + Unruh}

To connect the Schwinger pair production with the Unruh effect, let us consider the charged particle as the two level system with mass $M$ in the ground state level and mass $M+m$ in the excited level. According to Eq.(\ref{SchwingePairCreation}), the probability of creation of particles in the excited level with mass $M+m$ and charge $q$  can be expressed in terms of the probability of creation of particles in the ground state with mass $M$ and charge $q$ with the the extra term:
\begin{eqnarray}
\Gamma^{\rm Schw}(M+m) =\Gamma^{\rm Schw}(M)   \exp{\left(- \frac{2\pi  Mm}{q\cal E} - \frac{\pi  m^2}{q\cal E} \right)}.
\label{Mm}
\end{eqnarray}

In the limit $m\ll M$, the extra term can be described  in terms of the temperature of the Unruh radiation in the accelerated frame: 
\begin{eqnarray}
  \Gamma^{\rm Unruh}(m)=   \exp{\left(- \frac{m}{T_{\rm U}}\right)} \,,
 \label{Unruh}
 \\
T_{\rm U} = \frac{a}{2\pi} \,\,,\,\, a=\frac{q\cal E}{M}\,.
 \label{UnruhT}
  \end{eqnarray}
Here $a$ is the acceleration experienced by charged particles with mass $M$ after their creation in electric field. The accelerated particles with mass $M$ play the role of the Unruh-deWitt detectors  in the Unruh vacuum. These detectors experience the transition from the state with mass $M$ to the excited state with mass $M+m$ with probability determined by the Unruh temperature $T_{\rm U} = a/2\pi$. 

At first glance the problem was solved since the Unruh temperature here has the conventional form without factor 2. I informed Alexey about this result and published the paper.\cite{Volovik2022} However, 
the further considerations demonstrated that this is not so simple. In particular, the question arises on what happens if $m$ is not small, i.e. what is the role of the $m^2$ term in Eq.(\ref{Mm})? Let us first consider this question. 

\section{Back reaction of detector in Unruh effect}

The equation (\ref{Mm}) suggests that the Unruh contribution should be written in the form
\begin{eqnarray}
  \Gamma^{\rm Unruh}(m)=   \exp{\left(- \frac{m}{T_{\rm U}} \left(1+ \frac{m}{2M}\right)\right)} \,,
 \label{UnruhCorrected}
  \end{eqnarray}
and the correction $m/2M$ describes the recoil (or back reaction) of the detector due to its finite mass $M$.\cite{Parentani1997,Reznik1998,Casadio1999,KharzeevTuchin2005} This could be similar to the back reaction of the black hole in Hawking radiation discussed by Parikh and Wilczek.\cite{Wilczek2000}

To get this correction let us consider the step by step transitions to the higher mass of the detector, i.e. to the higher level $n$ of excitation of the detector.\cite{Reznik1998,Casadio1999} We consider $N$ steps, each with $\Delta m=m/N$. At each step the mass of the detector increases, $M_n=M+(n-1)\Delta m$ and thus the acceleration and the Unruh
temperature decrease correspondingly:
\begin{eqnarray}
 M_n=M+(n-1)\Delta m \,\,,\,\,  \Delta m=\frac{m}{N}\,,
  \label{steps}
  \\
  a_n=\frac{q{\cal E}}{M_n} \,\,,\,\, T_{{\rm U}n}=\frac{a_n}{2\pi}\,.
 \label{steps2}
 \end{eqnarray}
 Then one obtains Eq.(\ref{UnruhCorrected}):
\begin{eqnarray}
  \Gamma^{\rm Unruh}(m)=\prod_n \exp{\left(- \frac{\Delta m}{T_{{\rm U}n}} \right)}=  \exp{\left(- \sum_{n=1}^N \frac{\Delta m}{T_{{\rm U}n}} \right)}\,,
 \label{Unruh2}
 \\
=   \exp{\left(- \frac{m}{T_{\rm U}} \left(1+ \frac{m}{2M}\right)\right)} \,.
 \label{Unruh3}
  \end{eqnarray}

 \section{Back reaction in Hawking radiation}
 
 The same step-wise procedure can be applied to the Hawking radiation, where the black hole mass $M$ decreases after each $\Delta \omega=\omega/N$ step of Hawking radiation which step by step raises the Hawking temperature.\
 \begin{eqnarray}
 M_n=M-(n-1)\Delta \omega \,\,,\,\,  \Delta \omega=\frac{\omega}{N}\,,
  \label{HawkinhSteps}
  \\
 T_{{\rm H}n}=\frac{1}{8\pi GM_n}\,.
 \label{HawkingSteps2}
 \end{eqnarray}
 This gives the Parikh-Wilczek result:\cite{Wilczek2000}
 \begin{eqnarray}
  \Gamma^{\rm Hawking}(\omega)=%\prod_n \exp{\left(- \frac{\Delta \omega}{T_{{\rm H}n}} \right)}=
    \exp{\left(- \sum_{n=1}^N \frac{\Delta \omega}{T_{{\rm H}n}} \right)}\,,
 \label{Hawking3}
 \\
=   \exp{\left(- \frac{\omega}{T_{\rm H}} \left(1- \frac{\omega}{2M}\right)\right)} \,\,,\,\, T_{\rm H}=\frac{1}{8\pi GM}\,.
 \label{Hawking4}
  \end{eqnarray}
  The result is similar to that in Eq.(\ref{Unruh3}) except for the opposite sign, since the mass of the black hole decreases with radiation, while the mass of the detector increases its excitation by acceleration.
  
   \section{Double Unruh temperature from Schwinger+Unruh}
   
   So, using the corrected equation for the Unruh effect in Eq.(\ref{Unruh3}) we obtained the following relation between the Schwinger pair creation and the Unruh effect:
\begin{eqnarray}
\Gamma^{\rm Schw}(M+m) =\Gamma^{\rm Schw}(M)  \, \Gamma^{\rm Unruh}(m) \,.
 \label{SchwingerUnruh2}
 \end{eqnarray}
 
 However, we did not take into account that we discussed the coherent process of co-tunneling. This means that the Unruh effect here is the coherent process which combines the Unruh effect experienced by the accelerated particle with negative charge (electron) and the Unruh effect experienced by its partner with negative charge (positron). Since we want to get the probability of the Schwinger creation of the pairs with the same masses $M+m$, the Unruh processes must occur simultaneously. This means that their probabilities must be multiplied. That is why the Eq.(\ref{SchwingerUnruh2}) must be rewritten in the following form:
 \begin{eqnarray}
\Gamma^{\rm Schw}(M+m) =\Gamma^{\rm Schw}(M)  \, \Gamma_+^{\rm Unruh}(m) \Gamma_-^{\rm Unruh}(m)\,,
 \label{SchwingerUnruhDouble}
 \end{eqnarray}
 where $\Gamma_+^{\rm Unruh}(m)$ and $\Gamma_-^{\rm Unruh}(m)$ are the excitations rates of the corresponding detectors. 
 
 Comparing the equation (\ref{SchwingerUnruhDouble}) with the equation (\ref{Mm}), one finds that each of the two processes is governed by the temperature $\tilde T_{\rm U}$, which is twice the Unruh temperature: 
 \begin{eqnarray}
   \Gamma_\pm^{\rm Unruh}(m)=   \exp{\left(- \frac{m}{\tilde T_{\rm U}} \left(1+ \frac{m}{2M}\right)\right)} \,,
 \label{UnruhCoherent}
 \\
 \tilde T_{\rm U} = \frac{a}{\pi}=2T_{\rm U} \,\,,\,\, a=\frac{q\cal E}{M}\,.
 \label{UnruhDoubleT}
 \end{eqnarray}
 So, we returned back to the factor 2 problem.  But now we got the new argument in favour of the factor 2 in the Unruh effect.  The combined Schwinger-Unruh process suggests that the factor 2 in the equation (\ref{UnruhDoubleT}) is natural, and thus we must reconsider the temperature of the pure Unruh effect.
 The condensed matter analogs, where we know both the infrared and ultraviolet limits,\cite{Volovik2023} can be useful for that.
  
  \section{Double Unruh temperature in accelerating superfluid}
  
  That the factor 2 does make sense in the Unruh effect can be supported by its superfluid analog, see Sec. 6.15 "Pair Creation by Accelerated Object and Unruh Effect" in the book \cite{Volovik1992}.
  The equation (6.67) therein gives the following creation rate of the fermionic quasiparticles with energy  $E_{\bf k}$ at $k_z=0$ in the superfluid moving with acceleration $a$:
\begin{eqnarray}
  w \sim \exp - \frac{2\pi u E_{\bf k}}{\hbar a} \,.
 \label{UnruhSuperfluid}
 \end{eqnarray}
 Here the limiting velocity $u$ plays here the role of the speed of light. 
 
 At first glance the equation (\ref{UnruhSuperfluid}) describes the analog of the Unruh effect with the conventional Unruh temperature $T_{\rm U}=a/2\pi$. 
 However, we must take into account, that in this process the fermions are created in pairs (quasiparticle + quasihole) with the total energy $E=2E_{\bf k}$. Moreover, the process of the creation of quasiparticles and the process of creation of the quasiholes are coherent. That is why the equation (\ref{UnruhSuperfluid}) must be written in the following form:
\begin{eqnarray}
  w \sim \exp - \frac{2E_{\bf k}}{\tilde T_{\rm U}}\,\,, \,\, \tilde T_{\rm U} = \frac{a}{\pi}=2T_{\rm U} \,.
 \label{DoubleUnruhSuperfluid}
 \end{eqnarray}
Again, coherence of processes plays an important role in temperature doubling. Let us recall how a similar temperature doubling occurs in the de Sitter Universe.\cite{Volovik2024}

  \section{Double Gibbons-Hawking temperature in de Sitter Universe}
  
According to Ref.\cite{Volovik2024} the comoving observer perceives the de Sitter environment as the thermal bath with temperature $T=H/\pi$. It is twice larger than the Gibbons-Hawking temperature\cite{GibbonsHawking1977} of the cosmological horizon, $T_{\rm GH}=H/2\pi$. 
The  temperature $T=H/\pi$ determines in particular the process of ionization of an atom in the de Sitter environment: the rate of ionization is $w\propto \exp(-\frac{\epsilon_0}{T})$, where $\epsilon_0$ is the ionization potential. Here the atom plays the role of the local Unruh-deWitt detector, which is excited in the de Sitter environment.  

The ionization process is local, i.e. it takes place well inside the cosmological horizon, and thus this
local temperature $T=H/\pi$ has no relation to the cosmological horizon. Nevertheless, there is the close relation between these two temperatures: the local temperature is twice larger than the Gibbons-Hawking temperature. The factor 2 acquires the definite physical meaning, when we consider in more detail the Gibbons-Hawking radiation from the cosmological horizon. 

Again the main role here is played by the analog of co-tunneling, since in the Gibbons-Hawking process, two particles are coherently created: one particle is created inside the horizon, while its partner is simultaneously created outside the horizon.\cite{Parikh2002}  If the de Sitter Universe behaves as the thermal bath with temperature $T=H/\pi$, then the rate of the coherent creation of two particles, each with energy $E$, is  $w\propto \exp(-\frac{2E}{T})$. However, the observer who uses the Unruh-DeWitt detector can detect only the particle created inside the horizon. For this observer the creation rate $w\propto \exp(-\frac{2E}{T})$ is perceived as 
\begin{equation}
w\propto  \exp\left(-\frac{E}{T/2}\right)=\exp\left(-\frac{E}{T_{\rm GH}}\right)\,.
\label{HawkingRate}
\end{equation}
That is why the observer perceives the Hawking radiation from the cosmological horizon as the thermal process with the Gibbons-Hawking temperature $T_{\rm GH}=T/2=H/2\pi$, while the real temperature of the de Sitter environment is twice larger.

On the other hand, in the local process of the ionization of an atom, only single particle (electron) is radiated from the atom. This process is fully determined by the local temperature of the de Sitter environment $T=H/\pi$, with the rate $w\propto \exp(-\frac{\epsilon_0}{T})$. That is why both the local process and the process related to the cosmological horizon are governed by the same temperature $T=H/\pi$. 

It is important that this temperature has the real physical meaning. It determines the thermodynamics of the de Sitter state.\cite{Volovik2024} It gives in particular the local entropy density of this state:
\begin{equation}
 s_{\rm dS}=\frac{3\pi}{4G}T \,.
\label{EntropyDensity}
\end{equation}
The entropy density is linear in temperature, which demonstrates that de Sitter thermal state experiences the analog of the Sommerfeld law in Fermi liquids.

Although this temperature $T$ of the de Sitter environment is twice the Gibbons-Hawking temperature, the holographic principle does work. The total entropy $S_{\rm H}$ of the region inside the cosmological horizon coincides with the entropy of the cosmological horizon suggested by Gibbons and Hawking, which is determined by the area $A$ of the horizon: 
\begin{equation}
S_{\rm H}= s_{\rm dS}V_{\rm H}=\frac{A}{4G}\,.
\label{HubbleEntropy}
\end{equation}
Here $V_{\rm H}$ is the Hubble volume. The holographic relation between the total entropy $S_{\rm H}$ and the area $A$ of the cosmological horizon also supports the doubling of the temperature.

  \section{Double Hawking temperature in black hole}

The doubling of the Unruh and de Sitter temperatures takes place due to coherence of the several processes -- the analog of the co-tunneling.\cite{Feigelman2005,Glazman2005} This may also have some connection with the 't Hooft proposal of the doubling of the temperature of the Hawking radiation from the black hole.\cite{Hooft2022,Hooft2023} In the 't Hooft scenario, the temperature of the black hole horizon is $T= 2T_{\rm H}= \frac{1}{4\pi GM_{\rm BH}}$. In this scenario the coherence is supported by the partner (the clone) of the black hole -- the mirror image of the black hole space-time. 

Although the existence of such clone is problematic, the idea looks reasonable. Instead of the clone, the coherence can be provided by the simultaneous creation of two particles: the particle outside the black hole horizon and its partner -- the hole created inside the horizon. Due to coherence of these two processes the physical temperature is twice the Hawking temperature. However, the external observer has no information about the physics inside the horizon and according to Eq.(\ref{HawkingRate}) perceives the radiation as thermal with the Hawking temperature.

  \section{Conclusion}
 \label{ConclusionSec}

 The main conclusion is that the factor 2 in the Unruh process is not wrong. It follows from coherence of different processes. Due to such processes the temperatures of all three effects (Unruh effect, Gibbons-Hawking radiation from the de Sitter cosmological horizon and Hawking radiation from the black hole horizon) acquire the extra factor 2 compared with their traditional values. 
 
 In the case of de Sitter, the double Gibbons-Hawking temperature coincides with the thermodynamic temperature of the de Sitter state, which in particular responsible for the ionization rate of an atom in the de Sitter environment. This temperature also determines the local entropy $s$ of the de Sitter, which being integrated over the Hubble volume $V$ reproduces the entropy of the cosmological horizon, $sV=A/4G$, where $A$ is the horizon area.
 
 In the case of the Unruh effect, the double Unruh temperature is supported by the analog of the Unruh effect in the accelerated superfluid liquid such as $^3$He-B. In the Unruh process, two Bogoliubov fermions (quasiparticle and quasihole, each with energy $E$), are created simultaneously. Since the two fermions are created in unison, such coherent process looks as thermal but with the factor 2 in the exponent, $e^{-2E/T}$. This is the reason why the temperature $T$ corresponding to this coherent process is twice the Unruh temperature, $T=2T_{\rm U}$, where $T_{\rm U}= \hbar a/2\pi$ and $a$ is the acceleration of the liquid. 
  
 In case of the black hole Hawking radiation, the double Hawking temperature emerges also due to the combination of the coherent processes. Such coherence is similar to that in the scenario suggested by 't Hooft, where the black hole interior is considered as a quantum clone of the exterior region, which leads to the doubling of the Hawking temperature.
 
 The coherence of the processes is also used for the consideration of the back reaction of the black hole to the Hawking radiation and the detector recoil to the Unruh effect. In both cases the back reaction is calculated using the coherent set of small steps. In each step the mass of the black hole experience the stepwise decreases due to Hawking radiation, while in the Unruh effect the detector is stepwise excited. The obtained back reaction has the same value in both processes, but opposite sign.
 
Finally, the connection between the Schwinger pair production in electric field and the Unruh effect is established. We considered the combined process in which in the first step the pair of particles with masses $M$ are created by Schwinger mechanism. Then the created particles with positive and negative charges are accelerated by electric fields and they play the role of two Unruh-deWitt detectors. If due to the acceleration the mass of each detector is increased by $m$, the total process is equivalent to the pure Schwinger effect of creation of particles with masses $M+m$ in Eq. (\ref{SchwingerUnruhDouble}):
\begin{eqnarray}
\Gamma^{\rm Schw}(M+m) =\Gamma^{\rm Schw}(M)  \, \Gamma_+^{\rm Unruh}(m) \Gamma_-^{\rm Unruh}(m)\,.
 \label{SchwingerUnruhConnection}
 \end{eqnarray}
 The back reaction of the detectors is included into the rates of the Unruh processes $\Gamma_{\pm}^{\rm Unruh}(m)$ in Eq.(\ref{UnruhCoherent}).


\begin{thebibliography}{15}

\bibitem{Akhmedov2006}
E.T. Akhmedov, V. Akhmedova and D. Singleton,
Hawking temperature in the tunneling picture,
Phys. Lett. B {\bf 642}, 124--128 (2006),

\bibitem{Akhmedov2007}
E.T. Akhmedov, V. Akhmedova, T. Pilling and D. Singleton,
Thermal radiation of various gravitational background,
Int. J. Mod. Phys. A {\bf 22} 1705 (2007). 

\bibitem{Akhmedov2008}
E.T. Akhmedov, T. Pilling and D. Singleton, 
Subtleties in the quasi-classical calculation of Hawking radiation, 
Int. J. Mod. Phys. D {\bf 17},  2453 (2008).

\bibitem{Akhmedova2008}
V. Akhmedova, T. Pilling, A. de Gill, and D. Singleton, 
Temporal contribution to gravitational WKB-like calculations,
Phys. Lett. B {\bf 666}, 269--271 (2008). 

\bibitem{Akhmedova2009}
V. Akhmedova, T. Pilling, A. de Gill, and D. Singleton, 
Comments on anomaly versus WKB/tunneling methods for calculating Unruh radiation,
 Phys. Lett. B {\bf 673}, 227--231 (2009). 

\bibitem{Gill2010}
A.  de Gill,
A WKB-like approach to Unruh radiation,
American Journal of Physics {\bf 78}, 685 (2010).

\bibitem{Volovik2009}
G.E. Volovik,  
Particle decay in de Sitter spacetime via quantum tunneling,
Pis'ma ZhETF {\bf 90}, 3--6 (2009); 
JETP Lett. {\bf 90}, 1--4 (2009);
arXiv:0905.4639 [gr-qc].

\bibitem{Volovik2024}
G.E. Volovik, 
Thermodynamics and decay of de Sitter vacuum,
Symmetry {\bf 16}, 763 (2024),
https://doi.org/10.3390/sym16060763

\bibitem{Schwinger1951}
J. Schwinger, 
On gauge invariance and vacuum polarization,
 Phys. Rev. {\bf 82}, 664--679 (1951).
 
 \bibitem{Schwinger1953}
J. Schwinger, 
The theory of quantized fields. V,
Phys. Rev. {\bf 93},  616 (1953).

\bibitem{Unruh1976}
W.G. Unruh, 
Notes on black-hole evaporation,
Phys. Rev. D {\bf 14}, 870 (1976).

\bibitem{Parentani1997}
R. Parentani and S. Massar,
Schwinger mechanism, Unruh effect, and production of accelerated black holes,
Phys. Rev. D {\bf 55}, 3603 {1997}.

\bibitem{KharzeevTuchin2005}
D. Kharzeev and K. Tuchin,
From color glass condensate to quark–gluon plasma through the event horizon,
Nuclear Physics A {\bf 753},  316--334 (2005).

\bibitem{Kempf2021}
Vivishek Sudhir, Nadine Stritzelberger and Achim Kempf,
Unruh effect of detectors with quantized center of mass,
Phys. Rev. D {\bf 103}, 105023 (2021).

\bibitem{Zabrodin2022}
M. Teslyk, O. Teslyk, L. Zadorozhna, L. Bravina and E. Zabrodin,
Unruh effect and information entropy approach,
Particles {\bf  5}, 157--170 (2022).

\bibitem{Volovik1999}
G.E. Volovik,  
Simulation of Painleve-Gullstrand black hole in thin $^3$He-A film,  
Pis'ma ZhETF {\bf 69}, 662 -- 668 (1999), JETP Lett.  {\bf 69}, 705 -- 713 (1999); 
gr-qc/9901077.

\bibitem{Wilczek2000}
M.K. Parikh and F. Wilczek, 
Hawking radiation as tunneling,
Phys. Rev. Lett. {\bf 85}, 5042 (2000).

\bibitem{Hooft2022} 
G. 't Hooft, 
Quantum clones inside black holes, 
Universe {\bf 8}, 537 (2022).

\bibitem{Hooft2023} 
Gerard ’t Hooft, 
How an exact discrete symmetry can preserve black hole information or Turning a black hole
inside out,
 J. Phys.: Conf. Ser. {\bf 2533}, 012015 (2023).

\bibitem{Volovik2022}
G.E. Volovik,
Particle creation: Schwinger + Unruh + Hawking,
Pis’ma v ZhETF {\bf 116}, 577-578 (2022),
JETP Lett. {\bf 116}, 595--599 (2022),
DOI: 10.1134/S0021364022601968,
arXiv:2206.02799.


\bibitem{Reznik1998}
B. Reznik,
Unruh effect with back reaction—A first-quantized treatment,
Phys. Rev. B {\bf 57}, 2403 (1998).

\bibitem{Casadio1999}
Roberto Casadio and Giovanni Venturi,
The accelerated observer with back-reaction effects,
Physics Letters A {\bf 252}, 109--114 (1999).

\bibitem{Volovik2023}
G.E. Volovik, 
Gravity through the prism of condensed matter physics,
Pis’ma v ZhETF {\bf 118}, 546--547 (2023),
JETP Lett. {\bf 118},  531--541 (2023),
arXiv:2307.14370 [cond-mat.other].

\bibitem{Volovik1992}
G.E. Volovik, 
Exotic properties of superfluid $^3$He, 
World Scientific, Singapore-New Jersey-London-Hong Kong, 1992.

\bibitem{GibbonsHawking1977}
G.W. Gibbons and S.W. Hawking,
Cosmological event horizons, thermodynamics, and particle creation,
Phys. Rev. D {\bf 15}, 2738 (1977).

\bibitem{Parikh2002} 
M.K. Parikh,
New coordinates for de Sitter space and de Sitter radiation,
Physics Letters B {\bf 546},  189--195 (2002).

\bibitem{Feigelman2005}
M.V. Feigel’man and A.S. Ioselevich,
Variable-range cotunneling and conductivity of a granular metal,
JETP Letters {\bf 81},  277--283  (2005).

\bibitem{Glazman2005}
L.I. Glazman and  M. Pustilnik,
Low-temperature transport through a quantum dot,
Lectures notes of the Les Houches Summer School 2004,
in:  "Nanophysics: Coherence and Transport," eds. Helene Bouchiat, Yuval Gefen, Sophie Guéron, Gilles Montambaux, Jean Dalibard. (Elsevier, 2005), pp. 427--478,
arXiv:cond-mat/0501007.








\end{thebibliography}
\end{document}